\documentclass[12pt]{article}
\usepackage{epsfig}
 
\begin{document}
 
\thispagestyle{empty}
\renewcommand{\thefootnote}{\fnsymbol{footnote}}
 
\begin{flushright}
{\small
SLAC--PUB--8009\\
November 1998\\}
\end{flushright}
 
\vspace{.8cm}
 
\begin{center}
{\bf\large
The Tau Lepton and the Search for New Elementary Particle Physics\footnote{Work
supported by
Department of Energy contract  DE--AC03--76SF00515.}}
 
\vspace{1cm}
 
M.~L.~Perl\\
Stanford Linear Accelerator Center, Stanford University,
Stanford, CA  94309\\
 
\medskip
 
\end{center}
 
\vspace{1cm}
 
\begin{center}
{\bf\large
Abstract }
\end{center}

\begin{quote}
This Fifth International WEIN Symposium is devoted to physics beyond
the standard model.  This talk is about tau lepton physics, but I begin
with the question: do we know how to find new physics in the world of
elementary particles?  This question is interwoven with the various tau
physics topics.  These topics are: searching for unexpected tau decay
modes; searching for additional tau decay mechanisms; radiative tau
decays; tau decay modes of the $W$, $B$, and $D$; decay of the $Z^0$
to tau pairs; searching for CP violation in tau decay; the tau neutrino,
dreams and odd ideas in tau physics; and tau research facilities in the
next decades.
\end{quote}
 
\vfill
 
\begin{center}
{\it Invited talk presented at}\\
{\it The Fifth International WEIN Symposium: A Conference on Physics Beyond the\\
Standard Model
}\\
{\it Santa Fe, New Mexico}\\
{\it June 14--June 21, 1998}\\
 
\end{center}
 
\newpage

 
 
%

\newpage

\tableofcontents

\newpage

\pagestyle{plain}

\section{Do we know how to find new physics?}
\label{sec:1}

Over the years in elementary particle physics and to some extent in nuclear
physics we have developed a set of rules as to the best ways to look for new
physics. These rules are based on a set of axioms, perhaps better called
{\it id\'{e}es fixes}.
I am probably as trapped as other physicists in these
{\it id\'{e}es fixes}, but it is still worthwhile to think about them.
Perhaps we can loosen our minds a little or at least the younger practitioners
of particle and nuclear physics can develop some skepticism about the wisdom
of the elder practitioners.  

\begin{itemize}
\item Look for more massive particles.  The {\it id\'{e}es fixes} are:
	\begin{itemize}
	\item More massive particles exist.
	\item Supersymmetric theory is relevant.
	\item Mass is a significant property.
	\item More massive particles can be produced with present technology.
	\end{itemize}

\item Look for unconventional interactions.  The {\it id\'{e}es fixes} are:
	\begin{itemize}
	\item Unconventional interactions exist.	
	\item Sufficient experimental sensitivity can be obtained with present technology.
	\end{itemize}

\item Compare measured interaction and decay parameters with theory. The
{\it id\'{e}es fixes} are:
	\begin{itemize}
	\item There is a deeper theory that will lead to deviations.
	\item Sufficient experimental sensitivity can be obtained with present technology.
	\end{itemize}

\item Look for odd particles and peculiar phenomena. The {\it id\'{e}es fixes} are:  
	\begin{itemize}
	\item Such particles or phenomena exist.
	\item Sufficient experimental sensitivity can be obtained with present technology.
	\end{itemize}
\end{itemize}

	What other ways are there for new physics searches?  I don't have any
startling new ideas about how to look for new physics, but I have two somewhat
different ideas.  One idea is to go back and look with new technology at old
and fully established laws of physics.  Thus almost everyone believes there
are no free elementary particles with fractional electric charge; quarks are
not free. But my colleagues and I are using a highly automated version of the
Millikan oil drop experiment to search for particles with fractional electric
charge.\cite{Mar},\cite{PerlAm}  Another contrary idea is that mass is not an 
important property of elementary particles, beyond the law of relativistic 
energy conservation that allows heavy particles to decay into light particles. 
I don't know how to do experiments specifically based on this speculation, but 
it does incline me to be content to experiment in the realms of smaller masses
and lower energies.


\section{Overview of tau physics.}
\label{sec:2}

Tau physics includes studying the properties of the tau lepton and the 
properties of the tau neutrino.  This paper is primarily about the tau lepton 
but I briefly discuss tau neutrino research.  The recent exciting and important 
experimental results from the Super-Kamiokanda experiment on neutrino 
oscillations and their possible implications for the tau neutrino are 
thoroughly reviewed and discussed in this Symposium.\cite{Kajita}

I will not talk about using the tau to study the physics of hadrons and 
quantum chromodynamics; the subject lies outside the scope of the Symposium. 
It is however a vast and fruitful research area.

\begin{figure}[t!]
\centerline{
\epsfig{figure=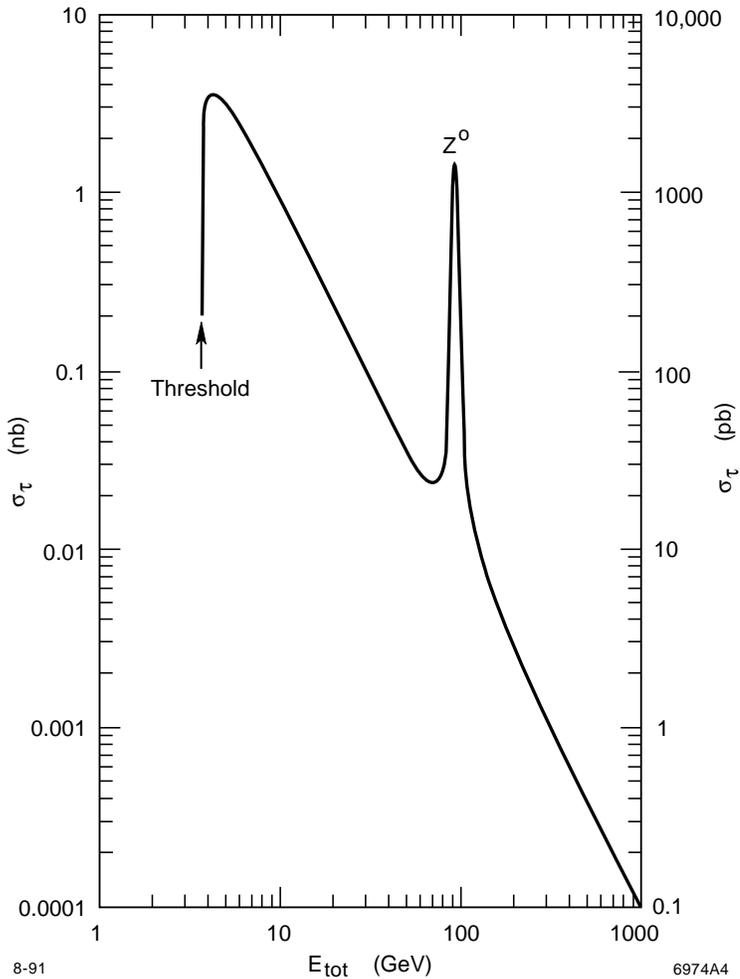,width=.6\textwidth}}
\caption{The energy dependence of the cross section for ${\tau}$ pair production.}
\label{fig:1}
\end{figure}

The energy dependence of the cross section for ${\tau}$ pair production
$$
e^+ + e^- \rightarrow {\tau}^+ + {\tau}^-
$$
is given in Fig.~\ref{fig:1}.  The discovery of the ${\tau}$ \cite{PerlPrix} and early 
${\tau}$ research was carried out in the energy region of  3 to 7~GeV using the 
first generation of electron-positron colliders.  The research continued with 
the next generation of colliders in the energy range from about 20~GeV to 
below the mass of the $Z^0$ (91~GeV).  The most recent period of ${\tau}$ 
research has
been carried out in two different energy regions.  Experimenters at LEP and the SLC have used the $Z^0$ decay
$$
	e^+ + e^- \rightarrow Z^0 \rightarrow {\tau}^+ + {\tau}^-.
$$
The other energy region, about 10~GeV, the region of $B$ meson pair production, 
has been a prolific source of ${\tau}$ research, first at the DORIS collider and
now at the CESR collider.  This 10~GeV region will be the main source of future 
${\tau}$ research as the CESR collider is upgraded and the $B$ factories go into
operation, see Sec.~\ref{sec:11}.

The ${\tau}$ has a great number of decay modes, summarized by the
Particle Data Group.\cite{Caso}  About 85\% of the time the ${\tau}$ decays
into one charged particle, neutral particles, and of course the tau neutrino,
${\nu_\tau}$. The major one-charged-particle, decay modes are the pure
leptonic modes: 
$$
\begin{array}{ll}
\tau^- \rightarrow \nu_\tau  + e^- + \nu_e&B = 17.8\%\\
\tau^- \rightarrow \nu_\tau  + \mu^- + \nu_{\mu}&B = 17.3\%\\
\end{array}
$$
In this paper I use ${\nu}$  to denote a neutrino or an antineutrino.
The branching fraction is denoted by $B$.

The major one charged particle, semileptonic, decay modes are:
$$
\begin{array}{ll}
\tau^- \rightarrow \nu_\tau  + \pi^-&B = 11.1\%\\
\tau^- \rightarrow \nu_\tau  + \rho^-&B = 25.3\%\\
\tau^- \rightarrow \nu_\tau  + \pi^- + 2\pi^0&B = 9.4\%\\
\end{array}
$$
Decay modes with a $K$ meson are much smaller because of Cabibbo suppression.

	In the decays with three charged particles, 9.6\% are of the form:
$$
	\tau^- \rightarrow \nu_\tau  + h^- +  h^+  +  h^- 
$$
where $h$ is a ${\pi}$ or a $K$.  The remaining 5.2\% have additional 
${\pi}^0$ or $K^0$ mesons.


\section{Searching for unexpected tau decay modes.}
\label{sec:3}

\begin{figure}
\centerline{
\epsfig{figure=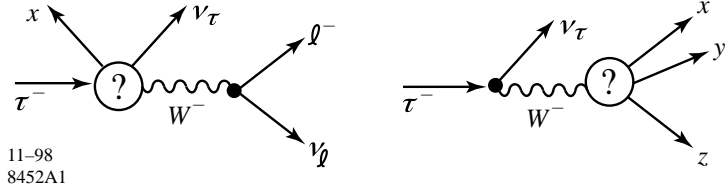}}
\caption{Unexpected decay modes of the tau.}
\label{fig:2}
\end{figure}

Some of us still dream about discovering unexpected decay modes of the tau,
decay modes that would lead to the discovery of new physics.
Figure~\ref{fig:2} shows two possibilities. In the upper diagram the
${\tau}$-$W$ vertex has unexpected behavior, another particle, $x$, is emitted.
The $W$ materializes into a standard pure leptonic final state.  In the
lower diagram, the ${\tau}$-$W$ vertex is standard, but the $W$ materializes
into an unknown $x$, $y$, $z$ final state.

But as yet there are no unexpected decay modes, all are explained by
conventional theory.  Thus for example:
\begin{itemize}
\item The total decay width of the tau is understood to better than 0.5\%.
Hence there are no undiscovered decay modes with branching fractions
greater than 0.5\%.
\item Lepton number violation such as ${\tau} \rightarrow {\mu}{\gamma}$
has not been observed. Experimenters using the CLEO detector\cite{Edwards}
have set the branching fraction upper limit
$B({\tau} \rightarrow {\mu}{\gamma}) < 3 \times 10^{-6}$.
Other lepton number violating modes have similar small upper limits. 
\item So far, expected modes with very small, branching fractions such as
${\tau}^- \rightarrow e^- e^+ e^- {\nu}_{\tau} {\nu}_e$
agree with theory. This decay mode has the branching fraction\cite{Alam}
$B = 3 \times 10^{-5}$.
\end{itemize}

{\it Yet I continue to ask myself: Are there other ways to search for
unexpected decay modes?}  


\section{Searching for additional tau decay mechanisms.}
\label{sec:4}

\begin{figure}
\centerline{
\epsfig{figure=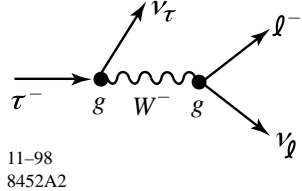}}
\caption{Standard decay route for tau.}
\label{fig:3}
\end{figure}

Perhaps there are no unexpected decay modes of the tau, but is there a 
tau decay route other than through a virtual $W$ with V-A coupling at
both vertices, Fig.~\ref{fig:3}?  The conventional theory for the
leptonic decays based on $W$ exchange is complete; hence it has been
the clearest way to search for other exchange mechanisms.  For the sake
of brevity, the discussion and examples in this section are limited to
the pure leptonic decays. 

Assuming Lorentz invariance, local interactions, no derivative couplings,
and lepton number conservation, the most general matrix element for these
decays consists of four different scalar interactions, four different
vector interactions, and two different tensor interactions. Each of these
interactions can have a different complex coupling constant. I use the
nomenclature of the classic paper of Fetscher.\cite{Fetscher}
The matrix elements, $m$, and coupling
constants, $g$, are denoted by:
$$
m^I_{LL^\prime}, \ m^I_{LR^\prime}, \ m^I_{RL^\prime}, \ m^I_{RR^\prime};
g^I_{LL^\prime}, \ g^I_{LR^\prime}, \ m^I_{RL^\prime}, \ m^I_{RR^\prime}  
$$
where $I$ $=$ $S$, $V$, $T$ for scalar, vector, or tensor interaction;
and the subscripts indicate whether the neutrinos are left-handed
($L$) or right-handed ($R$).

The matrix element for the V-A interaction of conventional theory is
denoted by $m^V_{LL}$, where $V$ means a vector interaction and the $LL$
means that both neutrinos are left handed.  This matrix element is
of course
\begin{equation}
m^V_{LL} = \left({\frac{g}{2 \sqrt{2}}}\right)^2 \frac{1}{M^2_W}
\left[\bar{u}({\nu}_{\tau}){\gamma}^{\mu} \left(1 - {\gamma}^5\right)u({\tau})\right]
\left[\bar{u}(\ell){\gamma}_{\mu}\left(1 - {\gamma}^5\right){\nu}({\nu}_{\ell})\right]
\end{equation}
$$
\frac{G_F}{\sqrt{2}} = \frac{g^2}{8M^2_W}
$$
The weak coupling constant is written simply as $g$ here and $G_F$ is the Fermi coupling constant.  

\begin{figure}
\centerline{
\epsfig{figure=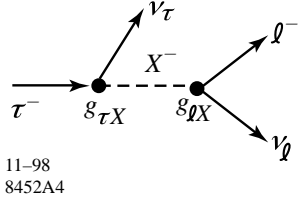}}
\caption{Possible scalar interaction in tau decay to leptonic maodes.}
\label{fig:4}
\end{figure}

As an example in the realm of new physics, a scalar interaction occurring
through the exchange of an unknown particle $X$, Fig.~\ref{fig:4},
might exist.  The matrix element for the leptonic decays would be:
\begin{equation}
m^S_{LL} = \left(\frac{g_{{\tau}{X}}g_{{\ell}{X}}}{8M^2_X}\right)
\left[\bar{u}({\nu}_{\tau})
\left(1 + {\gamma}^5)u({\tau}\right)\right]
\left[\bar{u}(\ell)\left(1 - {\gamma}^5\right)
{\nu}({\nu}_{\ell})\right]
\end{equation}
The only constraint on $m^S_{LL}$ is a constraint on an additive
combination of $\left|m^S_{LL}\right|^2$ and 
$\left|m^V_{LL}\right|^2$,\cite{Pich} hence a small amount of this scalar
interaction is certainly possible.  In Table~\ref{tab:CLlimits} I summarize
the present experimental limits on the possible ${\tau}$ coupling
constants and compare these limits with the tighter limits on the
${\mu}$ coupling constants.\cite{Pich}

\begin{figure}[t!]
\centerline{
\epsfig{figure=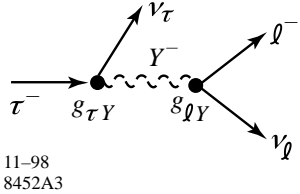}}
\caption{Possible additional vector interaction in tau decay to leptonic modes.}
\label{fig:5}
\end{figure}
As an example of how limits are set, consider a small second vector
interaction occurring thorough an unknown new particle $Y$, see
Fig.~\ref{fig:5}. The matrix element would be:
$$
m^{V'}_{LL} = \left(\frac{g_{{\tau}Y}g_{{\ell}Y}}{8M^2_Y}\right)
\left[\bar{u}({\nu}_{\tau}){\gamma}^{\mu} \left(1 - {\gamma}^5\right)u({\tau})\right]
\left[\bar{u}(\ell){\gamma}_{\mu}\left(1 - {\gamma}^5\right){\nu}({\nu}_{\ell})\right]
$$
From measurements of the ${\tau}$ lifetime and the leptonic
branching fractions, we can calculate the leptonic widths, and
compare them with standard V-A theory.  This comparison limits
$m^{V'}_{LL}$ as follows:
$$
\left|\frac{m^{V'}_{LL}}{m^V_{LL}}\right| < 5 \times 10^{-3}
$$
Hence there are constraints on the coupling constants in
Fig.~\ref{fig:5} and on the mass of the $Y$.

More generally, as shown in Table~\ref{tab:CLlimits}, {\it no deviations
from conventional physics have been found in the ${\tau}$ leptonic
decays}. But the limits are not nearly as tight as they are for the
corresponding coupling constants of the ${\mu}$. 

\begin{table}[t!]
\begin{center}
\caption{
\label{tab:CLlimits}
Comparison of 90\% CL limits on $\left|g^I\right|$'s from measurements of
${\mu} \rightarrow e {\nu}_e {\nu}_{\mu}$ with combined measurements
of ${\tau} \rightarrow e {\nu}_e {\nu}_{\tau}$  and
${\tau} \rightarrow {\mu} {\nu}_{\mu} {\nu}_{\tau}$.{\protect\cite{Pich}}
Because of the normalization convention some $\left|g^I\right|$'s have a
maximum value of 1, others have a maximum value of 2.{\protect\cite{Pich}}
In the table the numbers 1, 2, and $1/\sqrt{3}$ are
normalization limits, indicating there is no measured limit.
In conventional V-A theory all  $g^I$'s are zero except for
$g^V_{LL} = 1$.} 
\vskip 1em
\renewcommand{\baselinestretch}{1.4}
\small\normalsize
\begin{tabular}{|c|l|l|}
\hline
 & ${\mu} \rightarrow e {\nu}_e {\nu}_{\mu}$ &
${\tau} \rightarrow e {\nu}_e {\nu}_{\tau}$\\ 
 & & ${\tau} \rightarrow {\mu} {\nu}_{\mu} {\nu}_{\tau}$\\
\hline
$\left|g^S_{RR}\right| $ & $< 0.066$ & $< 0.57$\\
$\left|g^S_{LR}\right|$ & $< 0.125$ & $< 0.70$\\
$\left|g^S_{RL}\right|$ & $< 0.424$ & $\le 2$\\
$\left|g^S_{LL}\right|$ & $< 0.55$  & $\le 2$\\
\hline
$\left|g^V_{RR}\right|$ & $< 0.033$ & $< 0.29$\\
$\left|g^V_{LR}\right|$ & $< 0.060$ & $< 0.35$\\
$\left|g^V_{RL}\right|$ & $< 0.110$ & $< 0.53$\\
$\left|g^V_{LL}\right|$ & $> 0.96$  & $\le 1$\\
\hline
$\left|g^T_{LR}\right|$ & $< 0.036$ & $< 0.20$\\
$\left|g^T_{RL}\right|$ & $< 0.122$ & $\le 1/\sqrt{3}$\\
\hline
\end{tabular}
\renewcommand{\baselinestretch}{1}
\small\normalsize
\end{center}
\end{table}

Future searches for additional ${\tau}$ decay mechanisms will take
two directions:
\begin{itemize}
\item We can and will continue to improve measurements of leptonic
decay parameters.
\item In addition we should and do use probes that are intrinsically
sensitive to small additional decay mechanism. I will discuss two
such probes: radiative leptonic decays and searches for CP violation
in ${\tau}$ decay.
\end{itemize}

{\it
The first direction may soon reach its experimental limits, not
because of statistics but because of measurement problems in large
particle detectors.  The second direction is newer and has further
to go; there is more room for experimental progress.}


\section{What can we learn from radiative tau decays?}
\label{sec:5}

The radiation of a photon from a reaction offers the advantage that
the photon has no final state strong interaction.  Therefore the
kinematic properties of the emitted photon directly reflect the inner
dynamics of the reaction.  There is, however, a disadvantage: photons
are also emitted by the well-known but uninteresting process of
bremsstrahlung from charged particles produced or annihilated in the
reaction. These bremsstrahlung photons can lead to a serious
background problem.

The radiative decays of the tau fall into two classes: the purely
leptonic decays
\begin{eqnarray}
{\tau}^- \rightarrow {\nu}_{\tau} + e^- + {\nu}_e + {\gamma} \label{eq:lepdecaya}\\ 
{\tau}^- \rightarrow {\nu}_{\tau} + {\mu}^- + {\nu}_{\mu} + {\gamma} \label{eq:lepdecayb}
\end{eqnarray}
and the semileptonic decays such as
\begin{eqnarray}
{\tau}^- \rightarrow {\nu}_{\tau} + {\pi}^- + {\gamma} \label{eq:semilepdecaya}\\ 
{\tau}^- \rightarrow {\nu}_{\tau} + {\rho}^- + {\gamma} \label{eq:semilepdecayb}.
\end{eqnarray}
The conventional theory of radiative leptonic decays of charged
leptons is well established.\cite{Lenard}--\cite{Kinoshita}
Therefore the leptonic decays, Eq.~\ref{eq:lepdecaya}--\ref{eq:lepdecayb}, are valuable in
looking for unexpected tau decay phenomena.  The semileptonic
decays, Eq.~\ref{eq:semilepdecaya}--\ref{eq:semilepdecayb}, will be valuable for the study of $W$-hadron
vertices.\cite{Decker} In this talk I will concentrate on
the radiative leptonic decays.

\begin{figure}[t!]
\centerline{
\epsfig{figure=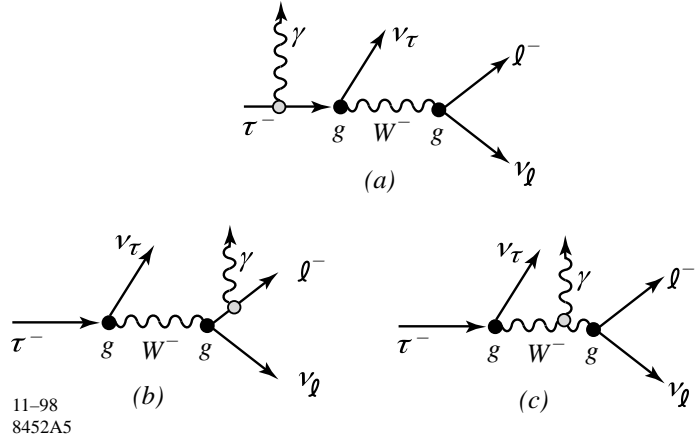}}
\caption{Diagrams for radiative, leptonic tau decays.}
\label{fig:6}
\end{figure}

The standard diagrams for radiative leptonic decay are given in
Fig.~\ref{fig:6}. The dominant contributions are from diagrams
a and b, radiation from the ${\tau}$ and radiation 
from the $e$ or ${\mu}$. Radiation from the $W$ is suppressed
by the very small factor $(M_{\tau}/M_W)^2$.\cite{Brunet},\cite{Bailin}

\begin{figure}
\centerline{
\epsfig{figure=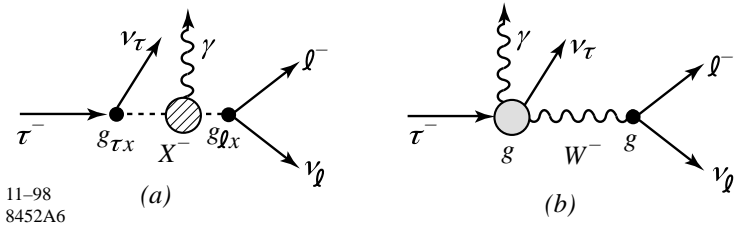}}
\caption{Possible types of anomalous, radiative, leptonic decays: (a) radiation from the exchange of particle $X$ and (b) anomalous radiation from the ${\tau}$-$W$ vertex.}
\label{fig:7}
\end{figure}

There are two types of anomalous, radiative, leptonic decay
processes that might exist, that we might dream about.  As shown in
Fig.~\ref{fig:7}a, there might be an unknown particle $X$ that is
exchanged in leptonic tau decays and that radiates.  Figure~\ref{fig:7}b
points out the possibility of the ${\tau}$-$W$ vertex having anomalous
photon radiation with a small branching ratio.  I have no model for
such anomalous vertex behavior but I think it is worth seeking.  

Returning to radiation from $X$ exchange, there are constraints on
the properties of $X$ and its coupling constants.  The mass of
$X$, $m_X$, must not be too large, otherwise the suppression factor
$(M_X/M_W)^2$ will be so small that we will not be able to detect
radiation from the $X$.  In terms of the coupling constants and 
$m_X$ this requires
\begin{equation}
\left|\frac{g_{{\tau}X}g_{{\ell}X}}{m^2_X}\right| << \left|\frac{g^2}{m^2_W}\right|.
\end{equation}

The radiative, muonic decay of the ${\tau}$ has been studied by
the OPAL experimenters,\cite{Alexander388} they find: 
$$
B ({\tau} \rightarrow {\nu}_{\tau} {\mu} {\nu}_{\mu} {\gamma}) =
(3.0 \pm 0.4 \pm 0.5) \times 10^{-3}
$$
for ${\gamma}$ energies above 20~MeV in the ${\tau}$ rest frame.
No anomalous behavior was found. CLEO experimenters\cite{Zhou}
are now studying the electronic and muonic radiative decays. 

\begin{figure}[t!]
\centerline{
\epsfig{figure=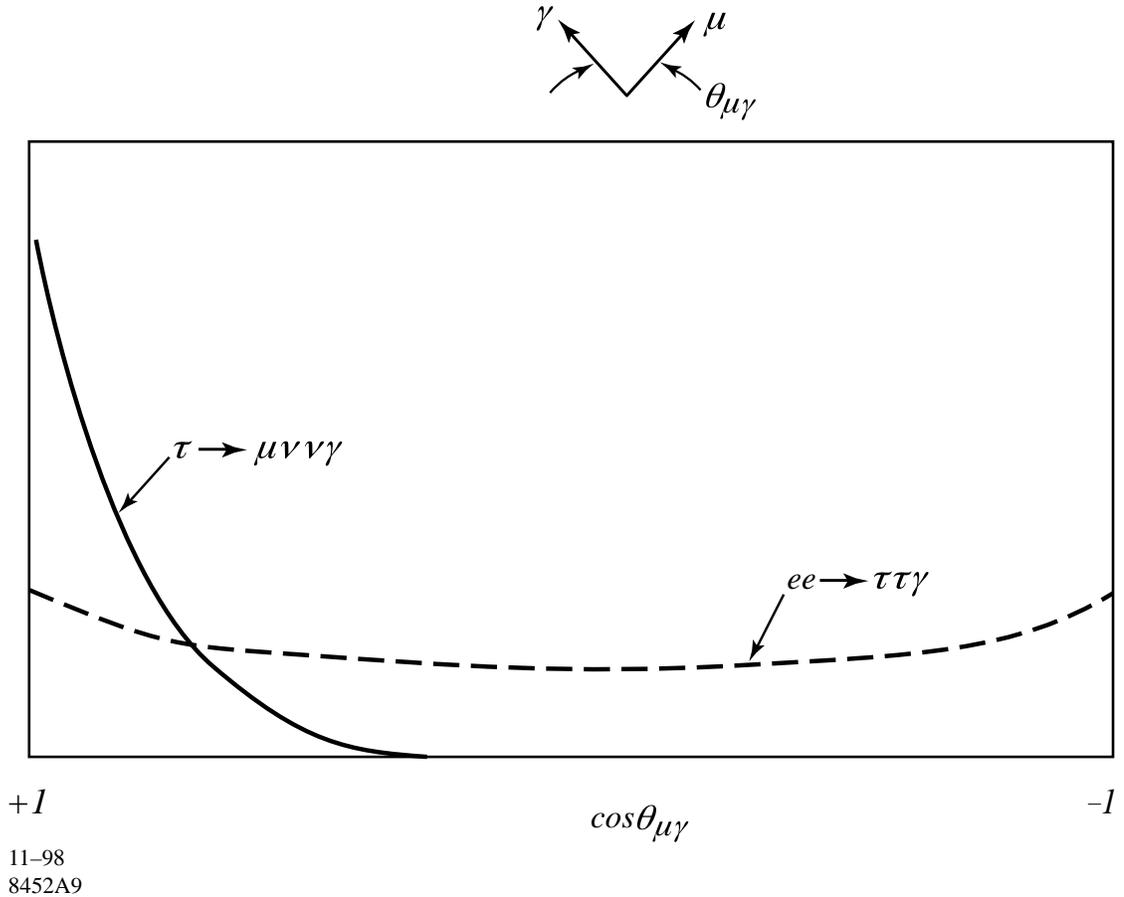,width=.9\textwidth}}
\caption{Schematic comparison of ${\tau} \rightarrow {\mu}{\nu}{\nu}{\gamma}$ signal and $e^+ e^- \rightarrow {\tau}{\tau}{\gamma}$ background.}
\label{fig:8}
\end{figure}

The major background in these studies is
\begin{equation}
e^+ + e^- \rightarrow {\tau}^+ + {\tau}^- + {\gamma},	
\end{equation}
the ${\gamma}$ being produced in the annihilation of the electron
or positron, or in the production of one of the ${\tau}$'s.  It is
an unfortunate background because, as shown schematically in
Fig.~\ref{fig:8}, it obscures what might be the most interesting
region for searches for anomalous radiation.  Consider the example
of radiative muonic decay.  We define ${\theta}_{{\mu}{\gamma}}$
as the angle in the laboratory frame between the ${\mu}$ and the
${\gamma}$.  The radiation from the standard diagrams, (a) and (b),
in Fig.~\ref{fig:6} peaks near ${\theta}_{{\mu}{\gamma}} = 0$.
Anomalous radiation might most easily be found when
${\theta}_{{\mu}{\gamma}} > 0$, but unless the anomalous radiation
has a relatively large branching fraction, it will be hidden by the
$e^+ e^- \rightarrow {\tau}^+ {\tau}^- {\gamma}$ radiation.

It will probably be difficult to substantially increase the precision
of these ${\tau}$ radiative decay studies, even with greatly increased
statistics. {\it We will have to find some new ways to explore the
radiative decays of the ${\tau}$.}


\section{\boldmath What can we learn from the tau decay modes of the $W$, $B$, and $D$?\unboldmath}
\label{sec:6}

\subsection{$W \rightarrow {\ell} + {\nu}_{\ell}$}
\label{subsec:6.1}

Since the $W$ mass is much larger than the lepton mass, we expect
the same branching fraction for all three leptons unless there is
a special connection between the $W$ and the ${\tau}$.  The Particle
Data Group\cite{Caso} gives the following branching fractions:
$$
B(W \rightarrow e + {\nu}_e) = 0.109 \pm 0.004
$$
$$
B(W \rightarrow {\mu} + {\nu}_{\mu}) = 0.102 \pm 0.005
$$
$$
B(W \rightarrow {\tau} + {\nu}_{\tau}) = 0.113 \pm 0.008
$$
Thus there is no apparent special connection between the ${\tau}$
and the $W$. {\it I don't think the precision of this comparison can be
improved substantially.}

\subsection{$D \rightarrow {\ell} + {\nu}_{\ell}, B \rightarrow {\ell} + {\nu}_{\ell}$}
\label{subsec:6.2}

The decay widths for 
$$
D^+ \rightarrow {\ell}^+ + {\nu}_{\ell}, \ \ 
D^+_S \rightarrow {\ell}^+ + {\nu}_{\ell}, \ \ 
B^+ \rightarrow {\ell}^+ + {\nu}_{\ell}
$$
are given by
\begin{equation}
{\Gamma} \left(M^+ \rightarrow {\nu}_{\ell} + {\ell}^+\right) =
\frac{G^2_F}{8 {\pi} {\hbar}} f^2_M V^2_{qq\prime}
m_M m^2_{\ell} \left[{1 - \left(\frac{m_{\ell}}{m_M}\right)^2}\right]^2
\label{eq:branfrac}
\end{equation}
Here $M$ is a $D$ or $B$ meson, $f_M$ is the meson decay constant
and $V_{qq\prime}$ is the CKM mixing matrix element. In
Table~\ref{table:branch}, I give the branching fractions using
Eq.~\ref{eq:branfrac}, assuming $f_M$ = 200~MeV.

\begin{table}[t!]
\begin{center}
\caption{
\label{table:branch}
Branching fraction for the purely leptonic decay,
$M \rightarrow {\ell} + {\nu}_{\ell}$ of a $D$ or $B$ meson. The
meson decay constant $f_M$ is assumed to equal 200~MeV and
$V_{qq\prime}$ is the assumed CKM mixing matrix element.}
\vskip 1em
\begin{tabular}{|l|l|l|l|l|}
\hline
\multicolumn{1}{|c|}{$M$}&\multicolumn{1}{c|}{$V_{qq'}$}&\multicolumn{1}{c|}{$e{\nu}$}&\multicolumn{1}{c|}{${\mu}{\nu}$}&\multicolumn{1}{c|}{${\tau}{\nu}$}\\
\hline
$D$& 0.221& $8 \times 10^{-9}$& $4 \times 10^{-4}$& $9 \times 10^{-4}$\\ 
$D_S$& 0.974& $7 \times 10^{-8}$& $3 \times 10^{-3}$& $3 \times 10^{-2}$\\ 
$B$& 0.003& $7 \times 10^{-12}$& $3 \times 10^{-7}$& $6 \times 10^{-5}$\\ 
\hline
\end{tabular}
\end{center}
\end{table}

\begin{table}[t!]
\begin{center}
\caption{
\label{table:branchtheory}
Comparison with theory of the measured branching fraction for the
purely leptonic decay, $M \rightarrow {\ell} + {\nu}_{\ell}$,
of a $D$ or $B$ meson. The meson decay constant is assumed to equal
200~MeV. Most measurements are upper limits.}
\vskip 1em
\begin{tabular}{|l|l|l|}
\hline
\multicolumn{1}{|c|}{Decay}&\multicolumn{1}{c|}{Branching Fraction, Data}&\multicolumn{1}{c|}{Branching Fraction, Theory}\\
&&\multicolumn{1}{c|}{($f_M=200$~MeV)}\\ 
\hline
$D \rightarrow {\mu}{\nu}$& $<7.2 \times 10^{-4}$\ & $4 \times 10^{-4}$\\
$D \rightarrow {\tau}{\nu}$&\ ?& $9 \times 10^{-4}$\\
$D_S \rightarrow {\mu}{\nu}$& $(4.0 \pm 2.1) \times 10^{-3}$& $3 \times 10^{-3}$\\
$D_S \rightarrow {\tau}{\nu}$& $(7 \pm 4) \times 10^{-2}$& $3 \times 10^{-2}$\\
$B \rightarrow e{\nu}$& $<1.5 \times 10^{-5}$& $7 \times 10^{-12}$\\
$B \rightarrow {\mu}{\nu}$& $<2.1 \times 10^{-5}$& $3 \times 10^{-7}$\\
$B \rightarrow {\tau}{\nu}$& $<5.7 \times 10^{-4}$& $6 \times 10^{-5}$\\
\hline
\end{tabular}
\end{center}
\end{table}

Table~\ref{table:branchtheory} presents a comparison with theory
of the measured branching fraction for the purely leptonic decay,
$M \rightarrow {\ell} + {\nu}_{\ell}$, of a $D$ or $B$ meson. Most
measurements are upper limits.  I have the following comments:
\begin{description}
\item[(a) ]Eventually we may know quite well the magnitude of $V_{qq\prime}$
from other measurements, but there is no independent way to measure
$f_M$.  Calculated values of $f_M$ can be used, particularly
if the calculation method is checked against a measured value
for a different $M$.  Therefore while we cannot hope to find new
physics by looking for small differences between calculated and
measured values of $B (M \rightarrow {\ell} {\nu}_{\ell})$ for
a particular decay mode, large differences between calculated and
measured values might indicate new physics.
\item[(b) ]One can get around a poorly known value of $f_M$ by using
the ratio of two different leptonic decays of the same $M$.
The accuracy of the calculated value of this ratio will of course
depend upon how well the $V_{qq\prime}$'s are known.
\item[(c) ]In the next decade there should be considerable progress
in measuring the ${\mu}$ and ${\tau}$ decay modes of the 
$D$ and $D_S$. The three charged particle decays of the ${\tau}$
will be most useful, but their use will require larger $D$ and $D_S$
data sets.
\item[(d) ] The measurement of the conventional theory values of the
branching fractions of $B$ decays in Table~\ref{table:branchtheory}
is beyond our present technology. Even the ${\tau}{\nu}$ mode is
very difficult because a very large number of $B$ pairs have to be
tagged.  {\it Still one can hope for new physics that drastically increases
these branching fractions}.
\end{description}

\subsection{$B \rightarrow {\tau}^+ + {\tau}^-$}
\label{subsec:6.3}

Another difficult but interesting direction for seeking an unexpected
connection between the $B$ and the ${\tau}$ is the search for the decays
$$
B^0_d \rightarrow {\tau}^+ + {\tau}^-
$$
$$
B^0_s \rightarrow {\tau}^+ + {\tau}^-
$$

Grossman {\it et al.}\cite{GrossmanRev} predict a branching fraction of
$10^{-6}$ to $10^{-8}$. Not easy to do. The present measured upper
limits are a few percent.


\section{\boldmath Decay of the $Z^0$ to tau pairs.\unboldmath}
\label{sec:7}

A tremendous amount of research work has been done on the $Z^0$ decays
\begin{eqnarray}
Z^0 \rightarrow e^+ + e^- \ \ \ \ {\Gamma} = 83.94 \pm 0.14~{\rm MeV} \\ 
Z^0 \rightarrow {\mu}^+ + {\mu}^- \ \ \ \ {\Gamma} = 83.84 \pm 0.20~{\rm MeV}\\ 
Z^0 \rightarrow {\tau}^+ + {\tau}^- \ \ \ \ {\Gamma} = 83.68 \pm 0.24~{\rm MeV}
\end{eqnarray}
Here ${\Gamma}$ is the decay width from the combined results of
measurements at LEP.\cite{Abbaneo} Thus in its overall
interaction with the $Z^0$, the ${\tau}$ behaves exactly like the
$e$ and ${\mu}$. There is no possibility of improving at LEP this
already fine precision because there will be no more LEP operation
at the $Z^0$. Perhaps some improvement may be made using the SLC
linear collider. {\it But we cannot expect much improvement.}

Hence it is important to look for more subtle deviations from expected
behavior in $Z^0 \rightarrow {\tau}^+ {\tau}^-$. An example is the
search for CP violation in $Z^0 \rightarrow {\tau}^+ {\tau}^-$.
No violation has been found.\cite{Wermes} The upper limit on any
such CP violation is usually given as an upper limit on a weak dipole
moment, $d^{weak}_{\tau}$:\cite{Wermes}
$$
\left|\mbox{Re}(d^{weak}_{\tau})\right| < 3.6 \times 10^{-18} \mbox{e cm}
$$
$$
\left|\mbox{Im}(d^{weak}_{\tau})\right| < 1.1 \times 10^{-17} \mbox{e cm}
$$
The level of $10^{-17}$ to $10^{-18}$ e cm does not seem particularly
small to me, since the equivalent size of a lepton is less than
$10^{-16}$ cm. 

{\it
In summary the extensive research on $Z^0 \rightarrow {\tau}^+ {\tau}^-$
leaves the feeling that the tau is simply an ordinary lepton.} 


\section{Searching for CP violation in tau decay.}
\label{sec:8}

We know that the violation of CP conservation occurs in the decays
of the $K$ mesons and there are strong reasons to believe that CP
violation also occurs in the decays of the $B$ mesons.  It is usually
believed that such CP violations can be explained by a theory that
concern only the quarks, even though no one
knows the correct theory.
But if the violation of CP conservation also occurred in lepton
decays, then we would need a more general and deeper theory of CP
violation. It is possible to search for CP violation in ${\mu}$
and ${\tau}$ decays, with the latter offering many more opportunities
for investigation.  

Unfortunately, even in ${\tau}$ decays it is difficult to search
for CP violation\cite{TsaiPh}--\cite{TsaiNu} for a number
of reasons:
\begin{itemize}
\item The ${\tau}$ has charge, therefore direct CP violation,
not mixing, is required.
\item The detection of CP violation in decay generally depends
upon the interaction of different phases.  One cannot search for CP
violation if the ${\tau}$ decays only through $W$ exchange. Therefore
the detection of CP violation in ${\tau}$ decays generally requires
the existence of a second exchange process.  This is good news in
the sense that the detection of CP violation in ${\tau}$ decay would
mean that new physics has been discovered. But it is bad news because
there may not be a second exchange process, or if there is such a
process its contribution to ${\tau}$ decay may be very small.
\item CP violation may be detected by using polarized ${\tau}$'s
or unpolarized ${\tau}$'s. In the latter case it is necessary to
look for differences between the angular distributions of the decay
products of the ${\tau}^+$ and the ${\tau}^-$.
\end{itemize}

\begin{figure}[t!]
\centerline{
\epsfig{figure=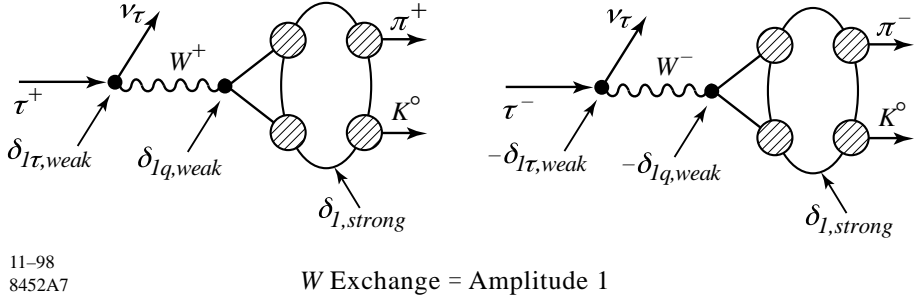}}
\caption{Diagrams for the decays ${\tau} \rightarrow {\nu}_{\tau} + {\pi}^{\pm} + K^0$ through $W$ exchange showing the phases.}
\label{fig:9}
\end{figure}

I use the decays
$$
{\tau}^\pm \rightarrow {\nu}_{\tau} + {\pi}^\pm + K^0
$$
to illustrate the requirements for detecting CP violation with
unpolarized taus. The $W$ exchange diagram is shown in Fig.~\ref{fig:9}
and this is called amplitude 1.  Under charge conjugation the weak
interaction phases at the $W$ vertices change sign, but the strong
interaction phase of the final states ${\pi}^\pm K^0$ does not change sign.

\begin{figure}[t!]
\centerline{
\epsfig{figure=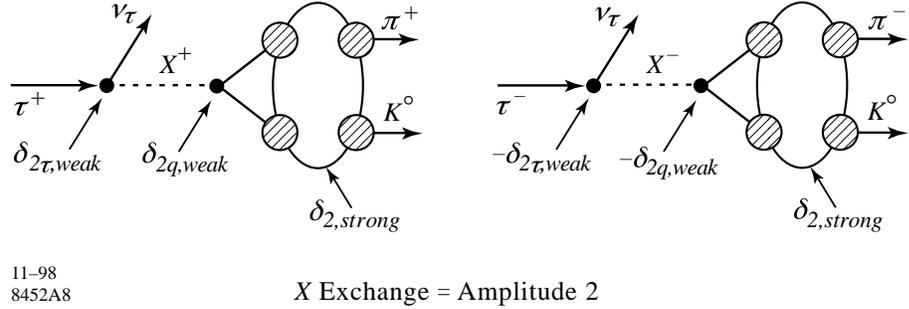}}
\caption{Diagrams for the decays ${\tau}^{\pm} \rightarrow {\nu}_{\tau} + {\pi}^{\pm} + K^0$ through $X$ exchange showing the phases.}
\label{fig:10}
\end{figure}

Figure~\ref{fig:10} shows the same decay occurring through an as yet
unknown exchange particle $X$, called amplitude 2.  The interference
of the $X$ exchange amplitude 2 with the $W$ exchange amplitude 1 can
give an asymmetry between the angular distribution of the two charge
states if amplitude 2 gives an angular distribution different from
amplitude 1.

The first search for CP violation in ${\tau}$ decay has been carried
out by CLEO experimenters\cite{Anderson} using mainly the decay
mode ${\pi}^+ K^0_S$ compared to ${\pi}^- K^0_S$, with the addition
of other decay modes.  No asymmetry was found in the angular
distributions.  This null result set the following upper limit on
CP violation in these tau decays:
$$
g_{{\rm CP violation}} \sin{\theta}_{{\rm CP violation}} < 1.7 g
$$
Here $g_{{\rm CP violation}}$ is the coupling constant for the a CP violating
effect, ${\theta}_{{\rm CP violation}}$ is a corresponding phase difference
in the amplitudes and $g$ is the standard weak coupling constant. Of
course this is not a surprising upper limit.  On the contrary we would
be surprised if $g_{{\rm CP violation}}$ were anywhere near the size of $g$. 

We should not be disappointed by the null result of this pioneer
experiment. There are many improvements that can be made to increase
the sensitivity. {\it With increased statistics in the next decade it will
be possible to probe to 1/5 to 1/10 of the
present limit.}

\pagebreak
CP violation involving leptons can also be sought in the reactions:
\cite{TsaiNu},\cite{GrossmanLett}
$$
\begin{array}{ll}
B \rightarrow {\tau} + {\nu}_{\tau} + {\rm hadrons},&
B \rightarrow {\mu} + {\nu}_{\mu} + {\rm hadrons}\\
\end{array}
$$
using ${\tau}$ or ${\mu}$ decay to measure polarization.


\section{\boldmath The tau neutrino, ${\nu}_{\tau}$.\unboldmath}
\label{sec:9}

The discovery of the tau and the recognition that a tau associated
neutrino, ${\nu}_{\tau}$, also existed led to four questions about
the properties of the ${\nu}_{\tau}$: does the ${\nu}_{\tau}$ have
conventional weak interactions; if the ${\nu}_{\tau}$ has a non-zero
mass, what is that mass; are there oscillations between the different
types of neutrinos, and in particular, does the ${\nu}_{\tau}$ partake
in those oscillations; and does the ${\nu}_{\tau}$ have any unusual
properties, for example is it stable?

Experimenters have labored for twenty years to try to answer these
questions, but we have had little success.  It is only in the last
year that we may have received positive evidence for a non-zero
${\nu}_{\tau}$ mass and for ${\nu}_{\tau}$  being involved in
neutrino oscillations.

\subsection{Interactions of the ${\nu}_{\tau}$?}
\label{subsec:9.1}

We are probably about to get evidence about the interactions of the
${\nu}_{\tau}$ with matter from experiments using a ${\nu}_{\tau}$ beam.
I have read that a few interactions of the ${\nu}_{\tau}$ have been
detected,\cite{Science} but there is no physics publication at present.  

If other neutrinos oscillate into the ${\nu}_{\tau}$, higher-energy
neutrino oscillation experiments may give us information about
${\nu}_{\tau}$ interactions. Of course for the present we assume
the interactions are conventional.

\subsection{${\nu}_{\tau}$ mass.}
\label{subsec:9.2}

The direct search for the ${\nu}_{\tau}$ mass using multiparticle
hadronic decays of the ${\tau}$ has been a frustrating business.
Recent measurements give an upper limit to the ${\nu}_{\tau}$ mass
in the range of 20 to 30~MeV/${\rm c}^2$.\cite{Ackerstaff}--\cite{Barate}
There are hopes that experimenters can probe down to the mass
range of several MeV/${\rm c}^2$ using this same method of
multiparticle hadronic decays with larger statistics.
{\it I am pessimistic.}

Of course the alternate way to explore the ${\nu}_{\tau}$ mass is
to use neutrino oscillation phenomena, if they exist for the ${\nu}_{\tau}$.

\subsection{Neutrino oscillations and the ${\nu}_{\tau}$.}
\label{subsec:9.3}

The most exciting particle physics news of this year is the research
by the experimenters using the Super-Kamiokande apparatus in Japan.\cite{Kajita}
The deficit of muon neutrinos from cosmic ray
interactions in the atmosphere can be interpreted as due to
oscillations between ${\nu}_{\mu}$'s and ${\nu}_{\tau}$'s. This
interpretation and the observations lead to
$$
\left|{\rm mass}^2 ({\nu}_{\tau}) - {\rm mass}^2 ({\nu}_{\mu})\right| = 
10^{-2} \ {\rm to}\  10^{-3} {\rm eV/c^2}
$$
Direct measurement of the ${\nu}_{\mu}$ mass gives an upper limit
of 170~keV/${\rm c}^2$; hence the ${\nu}_{\tau}$ mass
would have at least the same upper limit.  If we think there is no
reason for ${\nu}_{\mu}$ and ${\nu}_{\tau}$ to have very close large
masses, then the ${\nu}_{\tau}$ mass is of the order of
0.1~eV/${\rm c}^2$ or less.

Continuing with the idea that these observations\cite{Kajita} can
be interpreted as due to oscillations between ${\nu}_{\mu}$'s and
${\nu}_{\tau}$'s, one also obtains the mixing angle:
$$
\sin^2{\left(2{\theta}\right)} > 0.8,
$$
an amazing result, the mixing being so close to its maximum possible
value.

It looks like tau neutrino physics research finally has the needed
tools. {\it We in the tau physics community look forward to verification
and broadening of the observations made by the Super-Kamiokande
experimenters, as more experiments are turned on in this field.}

\subsection{Unconventional ${\nu}_{\tau}$ properties?}
\label{subsec:9.4}

Are there other unconventional properties of the ${\nu}_{\tau}$
besides the possibility that it can oscillate into other neutrinos?
For example, is the ${\nu}_{\tau}$ stable?

We cannot make much progress in answering such questions until we can
directly study the interactions of the ${\nu}_{\tau}$, see Sec.~\ref{subsec:9.1}.
But astrophysical calculations and observations have taught us
some things. Other things about the ${\nu}_{\tau}$ have been deduced
from terrestrial research using beam dump experiments. The Particle
Data Group\cite{Caso} lists lower limits on the lifetime of
the ${\nu}_{\tau}$ and upper limits on the magnetic moment, the
electric dipole moment, and the electric charge.  The review by
Gentile and Pohl\cite{Gentile} also discusses these limits.
{\it As you probably know no anomalies have been found in these quantities.}


\section{Dreams and odd ideas in tau research.}
\label{sec:10}

I conclude the physics discussions with three examples of the many
odd ideas and dreams of the tau research community.

\subsection{Tau magnetic moment.}
\label{subsec:10.1}

Recall that the magnetic moment of a ${\tau}$ is expected to be\cite{Gentile}
$$
{\mu}_{\tau} = \frac{g_{\tau}eh}{4{\pi}m_{\tau}}, \ \ \ 
g_{\tau} = 2 \left[1 + a_{\tau}\right], \ \ \ 
a_{\tau} = \frac{{\alpha}}{2{\pi}} + O\left(\frac{\alpha}{\pi}\right)^2 + ...
$$
The Schwinger term, ${\alpha}/2{\pi}$, in $a_{\tau}$ has the value 
$$
a_{\tau} = 0.0012.
$$
It would be very nice to measure ${\mu}_{\tau}$ with enough
precision to check this, as it was checked for the $e$ and the
${\mu}$ years ago.  {\it At present such precision is a dream.}  The
best that has been done so far is to use the decay
$Z^0 \rightarrow {\tau}^+ + {\tau}^- + {\gamma}$.  Acciarri {\it et al}.\cite{Acciarri}
found the limits
$$
a_{\tau} = 0.004 \pm 0.027 \pm  0.023;
$$
thus the limits are an order of magnitude larger than the expected value.

\subsection{Is ${\tau}$ decay exponential?}
\label{subsec:10.2}

Alexander {\it et al}.\cite{Alexander368} have examined the distribution
of the decay time of individual ${\tau}$'s to test if the decay
distribution is exponential.  The ${\tau}$, being a relatively
heavy elementary particle, is a good specimen for such a test.
They find that ${\tau}$ decay is quite ordinary; it is exponential
to better than 10\%.

\subsection{A ${\tau}^+$-${\tau}^-$ atom and a ${\tau}^-$ nucleus atom?}
\label{subsec:10.3}

One of my dreams is to make ${\tau}^+$-${\tau}^-$ atoms, in analogy
to $e^+$-$e^-$ atoms, positronium.  This can be done just below
${\tau}$ pair threshold at a Tau-Charm factory.\cite{PerlSecond}--\cite{PerlFifth} It would be a {\it tour de force}; perhaps it might be
a way to look for new physics, although I admit that is a very long
shot.  One can also think about making a ${\tau}$-nucleus atom, in
analogy to mu-mesic atoms; again a long shot.\cite{PerlSecond}


\section{Tau research facilities in the next decade.}
\label{sec:11}

\subsection{Present tau research facilities.}
\label{subsec:11.1}

The major existing collections of tau data are:
\begin{itemize}
\item The CLEO experimenters have about $10^7$ ${\tau}$ pairs collected
at 10~GeV; the analysis of this data is continuing.
\item The ALEPH, DELPHI, L3, OPAL and SLD have each collected about
$10^5$ or more ${\tau}$ pairs from $Z^0$ decays.  The analysis is
mostly completed; the SLD experimenters may collect more data. Note
that the efficiency of using ${\tau}$ pairs in analysis is 2 to 5
times higher at 91~GeV, the $Z^0$ energy, compared to 10~GeV.
\item The BEPC experimenters have about $10^5$ ${\tau}$ pairs
collected in the 4~GeV region.
\end{itemize}

\subsection{Future tau research facilities.}
\label{subsec:11.2}

In the next few years three very high luminosity $B$ factories will
begin full operation at about 10~GeV; these are: the symmetric CLEO
III-CESR III facility in New York, the asymmetric Belle-KEKB facility
in Japan, and the asymmetric BABAR-PEPII facility in California.
Each of these facilities expect initial luminosities in full operation
of $1 \times 10^{33}$ cm$^{-2}$s$^{-1}$; this will yield
$1.5 \times 10^7$ ${\tau}$ pairs/year.  (I assume $1.5 \times 10^7$
seconds/year for data acquisition.)  Once the design luminosities
for these facilities of $3 \times 10^{33}$ cm$^{-2}$s$^{-1}$ is
reached, each facility will yield $5 \times 10^7$ ${\tau}$ pairs/year.
Another factor of 3 increase in ${\tau}$ pairs/year will be obtained
if the perhaps dream luminosity of $1 \times 10^{34}$ cm$^{-2}$s$^{-1}$
is obtained.

Some final remarks.  The upgrade of BEPC will yield about 3 to
$5 \times 10^5$ ${\tau}$ pairs/year.  Meanwhile there will be continual
production of ${\tau}$ pairs at the LEP experiments for the next few
years.  Since the energy will be above 180~GeV, the number of ${\tau}$
pairs will be relatively small.  Still it will be interesting to see if
there is any anomaly in the cross section for
$e^+ + e^- \rightarrow {\tau}^+ + {\tau}^-$
at this very high energy.

In the past decade there has been much work on the concept of a
Tau-Charm factory: a high luminosity electron positron collider designed
to operate primarily in the 3 to 4~GeV region.\cite{Kirkby}--\cite{PerlFifth}
The facility would include a detector designed specially for
the study of tau physics and charm physics.  At present there are no
firm plans to build a Tau-Charm factory. 

In conclusion:
\begin{itemize}
\item {\it In the next decade the amount of accumulated ${\tau}$
decay data will increase by a factor of ten and perhaps eventually
by another factor of ten!}  
\item {\it Full and fruitful use of this increased statistics will
require substantial improvements in increasing the precision and
decreasing the bias of existing particle detectors.}
\item {\it Tau physics has a marvelous future.}
\end{itemize}

\section*{Acknowledgments}

I am grateful to the organizers and staff of this Fifth International
WEIN Symposium for having organized such a stimulating meeting and for
having given me the opportunity to present this paper.

In this paper I have depended a great deal on some recent reviews of
tau physics:
Pich,\cite{Pich} Gentile and Pohl,\cite{Gentile} and Weinstein
and Stroynowski.\cite{Weinstein}  I recommend these reviews to the
reader.



\begin{thebibliography}{99}
\bibitem{Zhou} This work is being carried out by my colleague Xiaofan Zhou and myself with other members of the CLEO Collaboration.
\end{thebibliography}
\end{document}